\documentclass[aps,prx,superscriptaddress,amsmath,amssymb,twocolumn,showpacs,floatfix,reprint]{revtex4-2}
\usepackage[colorlinks=true, urlcolor=blue, linkcolor=blue, citecolor=blue, pdftex]{hyperref}
\usepackage[capitalise]{cleveref}
\usepackage{xr}
\usepackage{multirow}
\usepackage[dvipsnames]{xcolor}
\usepackage[utf8]{inputenc}
\usepackage{braket}
\usepackage{graphicx}

\begin{document}
\title{Transformer Neural-Network Quantum States for lattice models\\ of spins and fermions: Application to the Ancilla Layer Model
}

\author{Riccardo Rende}
\thanks{These authors contributed equally. Correspondence should be addressed to rrende@flatironinstitute.org and onikolaienko@g.harvard.edu}
\affiliation{Center for Computational Quantum Physics, Flatiron Institute, 162 5th Avenue, New York, NY 10010}

\author{Alexander Nikolaenko}
\thanks{These authors contributed equally. Correspondence should be addressed to rrende@flatironinstitute.org and onikolaienko@g.harvard.edu}
\affiliation{Department of Physics, Harvard University, Cambridge MA 02138, USA}

\author{Luciano Loris Viteritti}
\affiliation{Institute of Physics, \'{E}cole Polytechnique F\'{e}d\'{e}rale de Lausanne (EPFL), CH-1015 Lausanne, Switzerland}

\author{Subir Sachdev}
\affiliation{Center for Computational Quantum Physics, Flatiron Institute, 162 5th Avenue, New York, NY 10010}
\affiliation{Department of Physics and Leinweber Institute for Theoretical Physics, Harvard University, Cambridge MA 02138, USA}

\author{Ya-Hui Zhang}
\affiliation{Department of Physics and Astronomy, Johns Hopkins University, Baltimore, Maryland 21218, USA}

\date{\today}

\begin{abstract}
   We introduce a variational wave function based on Neural-Network Quantum States (NQS) to study lattice systems whose local Hilbert space contains both spin and fermionic degrees of freedom. Our approach is based on the use of the Transformer architecture, which can naturally handle composite local Hilbert spaces through a tokenization procedure closely inspired by techniques from natural language processing. The key methodological innovation is that the neural network predicts a set of fermionic orbitals that depend on the spin configuration in a backflow-inspired manner. We apply the method to the one-dimensional Ancilla Layer Model, consisting of a chain of mobile spin-$1/2$ fermions coupled to a two-leg spin-$1/2$ ladder. For open boundary conditions, we achieve excellent quantitative agreement with Density Matrix Renormalization Group (DMRG) results across the full range of parameters considered. We find  a phase in which the chain forms an effectively decoupled Luttinger liquid (LL), and a LL* phase with a distinct Fermi wavevector in which the mobile fermions are Kondo screened by one leg of the ladder, while the other leg forms the critical Bethe spin liquid. The LL* is the analog of the FL* phase in two dimensions proposed to describe the cuprate pseudogap. We also find a Luther-Emery (LE) phase, where the LL* state becomes unstable toward the formation of a spin gap. The Transformer Ansatz maintains comparable accuracy for periodic boundary conditions, where tensor-network methods are computationally more demanding. Moreover, we show that the approach extends naturally to fully periodic two-dimensional clusters, a regime beyond the reach of current tensor-network methods. Together, these findings establish Transformer-based NQS as an accurate and scalable variational framework for correlated lattice systems with composite local Hilbert spaces and highlight their potential for studying higher-dimensional models where boundary effects and heterogeneous local structures pose significant challenges.
\end{abstract}

\maketitle

\section{Introduction}
\label{sec:intro}

Variational representations of quantum many-body wave functions based on neural networks have emerged as a powerful complement to traditional numerical approaches~\cite{white1992, Foulkes2001, sandvik2010, Verstraete2008, Schollwock2011}. Neural-Network Quantum States (NQS)~\cite{carleo2017,lange2024} provide flexible and systematically improvable approximations to many-body quantum states and have been successfully applied to a broad range of spin and fermionic lattice models~\cite{rende2024stochastic,lange2024,viteritti2026,gu2025solvinghubbardmodelneural,roth2025,chen2024empowering}. A key ingredient underlying this success is the ability of neural architectures to efficiently represent high-dimensional functions living in exponentially large Hilbert spaces with polynomial resources. However, in many physically relevant settings, the structure of the local Hilbert space itself poses a nontrivial challenge to this paradigm. In particular, a wide class of quantum lattice models is characterized by \emph{composite local Hilbert spaces}, in which multiple local degrees of freedom, such as fermionic occupations, localized spins, orbital indices, or auxiliary ancilla variables, coexist on each lattice site~\cite{hubbard1963, kanamori1963, doniach1977, georges2013, hirschfeld1989, capone2009}. Incorporating this internal structure into variational ansatze is typically achieved through model-specific parametrizations~\cite{nomura2020, marino2025, piccioni2025, favata2025, mahajan2025}. While effective in targeted applications, such constructions can limit flexibility and transferability, especially as the number or nature of local degrees of freedom increases.

\begin{figure*}[t]
    \begin{center}
\centerline{\includegraphics[width=2\columnwidth]{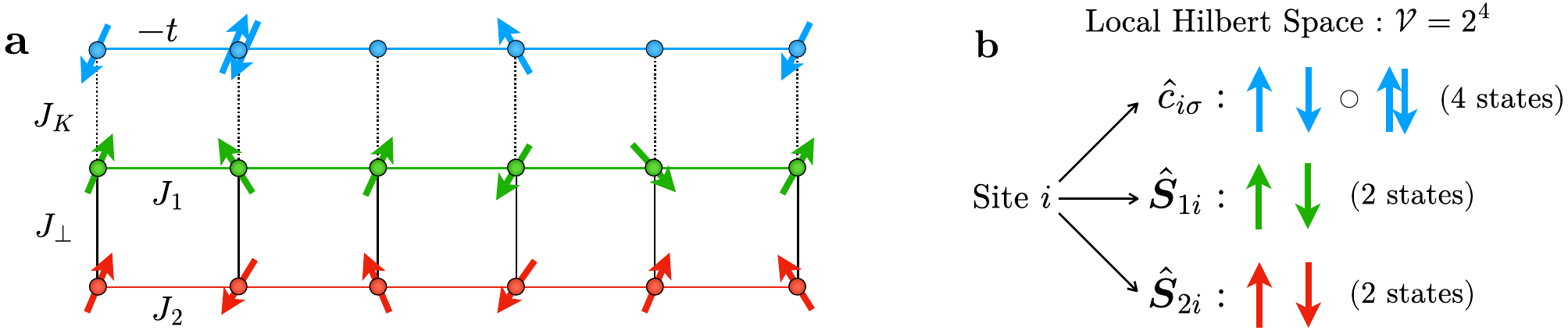}}
        \caption{\label{fig:alm_model} \textbf{Panel a:} Schematic representation of the one-dimensional Ancilla Layer Model [see \cref{eq:hamiltonian}]. A chain of itinerant spinful fermions (blue arrows) of density $1-\delta$ is locally coupled via a Kondo exchange $J_K$ to the first leg of a spin-$\tfrac12$ ladder (green arrows). The two ancillary spin layers are coupled along the chain direction by Heisenberg interactions $J_1$ and $J_2$, respectively, and across the rungs by an interlayer coupling $J_\perp$. Each lattice site thus hosts both fermionic and spin degrees of freedom, resulting in a composite local Hilbert space. \textbf{Panel b:} Each lattice site hosts two ancillary spin-$\tfrac12$ variables and a spinful fermionic occupation, giving a composite local Hilbert space of dimension $\mathcal{V}=2^4$.}
    \end{center}
\end{figure*}

Transformer neural networks offer a natural framework to address this challenge~\cite{vaswani2017attention,wolf2020transformers,surveytransformers}. Originally developed for sequence modeling in natural language processing, these architectures represent data as collections of interacting tokens and capture correlations through self-attention mechanisms~\cite{vaswani2017attention}. This structure enables the modeling of long-range and heterogeneous dependencies without imposing locality constraints. Recent works have demonstrated that Transformer-based wave functions can achieve state-of-the-art accuracy for quantum many-body systems for both spin~\cite{viteritti2023prl, rende2024stochastic,lange2024,sprague2024, viteritti2025prb, viteritti2026,viteritti2026quantumspinglasstwodimensional,rende2025foundations} and fermionic~\cite{vonglehn2023, gu2025solvinghubbardmodelneural,sharma2025, ma2025,Shang2025QiankunNet, Teng2025} Hamiltonians. However, their ability to systematically and efficiently encode both degrees of freedom within a unified architecture has not yet been fully exploited.

In this work, we introduce a Transformer-based NQS for lattice models with composite local Hilbert spaces. We perform an explicit \emph{tokenization}~\cite{jm3} of the local variables, mapping different degrees of freedom on each lattice site to distinct tokens to construct the input sequence. This representation, inspired by techniques from natural language processing, enables the network to learn both intra- and inter-site correlations within a single unified architecture. The novelty of our approach is to exploit the output of the neural network to construct a set of backflow fermionic orbitals that depend on the spin variables. As a result, this variational Ansatz can be extended straightforwardly to more complicated settings, such as multi-band fermion models with multiple orbitals per site~\cite{georges1996, capone2009, georges2013}.

We apply this approach to the one-dimensional Ancilla Layer Model (ALM), which consists of a fermionic chain locally coupled to a two-leg spin-$\tfrac12$ ladder through a Kondo-type exchange~\cite{Boulder25}. The Hamiltonian reads
\begin{equation}\label{eq:hamiltonian}
\begin{aligned}
    \hat{H} &=\;  
    -t \sum_{i} \sum_{\sigma=\uparrow,\downarrow} 
        \!\!\left( \hat{c}^\dagger_{i,\sigma} \hat{c}_{i+1,\sigma} + \mathrm{h.c.} \right) \\
     &+ J_1 \sum_{i} 
        \hat{\boldsymbol{S}}_{i,1} \cdot \hat{\boldsymbol{S}}_{i+1,1}
    + J_2 \sum_{}
        \hat{\boldsymbol{S}}_{i,2} \cdot \hat{\boldsymbol{S}}_{i+1,2} \\
    &+ \frac{J_K}{2} \!\sum_{i} \hat{\boldsymbol{S}}_{i,1} \cdot \!\!\!\!\!
    \sum_{\sigma,\sigma'=\uparrow,\downarrow}\!\!\!\!
    \hat{c}_{i,\sigma}^\dagger \,\boldsymbol{\tau}_{\sigma\sigma'}\, \hat{c}_{i,\sigma'} + J_\perp \!\!\sum_{i}
        \hat{\boldsymbol{S}}_{i,1} \cdot \hat{\boldsymbol{S}}_{i,2} \ .
\end{aligned}
\end{equation}

Here $\hat{c}_{i,\sigma}^\dagger$ ($\hat{c}_{i,\sigma}$) is the fermionic creation (annihilation) operator on site $i$, $\sigma$ denotes the spin $(\uparrow$ or $\downarrow)$ , and
$\boldsymbol{\tau}=(\tau^x,\tau^y,\tau^z)$ denote the Pauli matrices acting in spin space. The operators $\hat{\boldsymbol{S}}_{i,1}$ and $\hat{\boldsymbol{S}}_{i,2}$ are spin-$\tfrac{1}{2}$ degrees of freedom on the two ancillary layers at site $i$. In what follows, we consider a one-dimensional chain of $N$ sites with either periodic (PBC) or open (OBC) boundary conditions. The total number of electrons is $N_e=N_{\uparrow}+N_{\downarrow}$, and we define the hole doping as $\delta = 1 - N_e/N$.

The Hamiltonian in \cref{eq:hamiltonian} contains different contributions; refer to \cref{fig:alm_model} for a pictorial representation.
The first term describes nearest-neighbor hopping along the chain with amplitude $t$. In what follows, we set the hopping amplitude to $t = 1$ and express all remaining couplings in units of $t$.
The second and third terms are nearest-neighbor Heisenberg interactions along the first and second ancillary spin layers, with antiferromagnetic couplings $J_1, J_2$, respectively.
The remaining terms represent the onsite Kondo exchange, with strength $J_K > 0$ between the electronic spin density and a localized spin $\hat{\boldsymbol{S}}_{i,1}$, and an antiferromagnetic interaction of strength $J_\perp$ between two ancillary spins $\hat{\boldsymbol{S}}_{i,1}$ and $\hat{\boldsymbol{S}}_{i,2}$ on each rung. 

An important point is that the phases of the ALM are expected to be the same as those that could be obtained by extending the single-band Hubbard model with short-range interactions at the same doping $\delta$; see Ref.~\cite{Boulder25} for a review. This can be seen explicitly
in the limit $J_\perp \rightarrow \infty$, where the ALM becomes a conventional Hubbard model with the on-site interaction $U\propto J_K^2/J_\perp$ \cite{Maria21,Mascot2022}. More generally, we are allowed to add ancilla degrees of freedom with a trivial gapped ground state because such degrees of freedom can be safely integrated out, leaving only additional short-range interactions between the mobile fermions; the bilayer antiferromagnet of the $\boldsymbol{S}_{1,2}$ is the simplest system that fulfills these requirements. In one dimension, there is also a resemblance of this structure to that in the `spin-gap proximity effect' \cite{Emery97}.

\begin{figure}[t]
    \begin{center}
\centerline{\includegraphics[width=0.95\columnwidth]{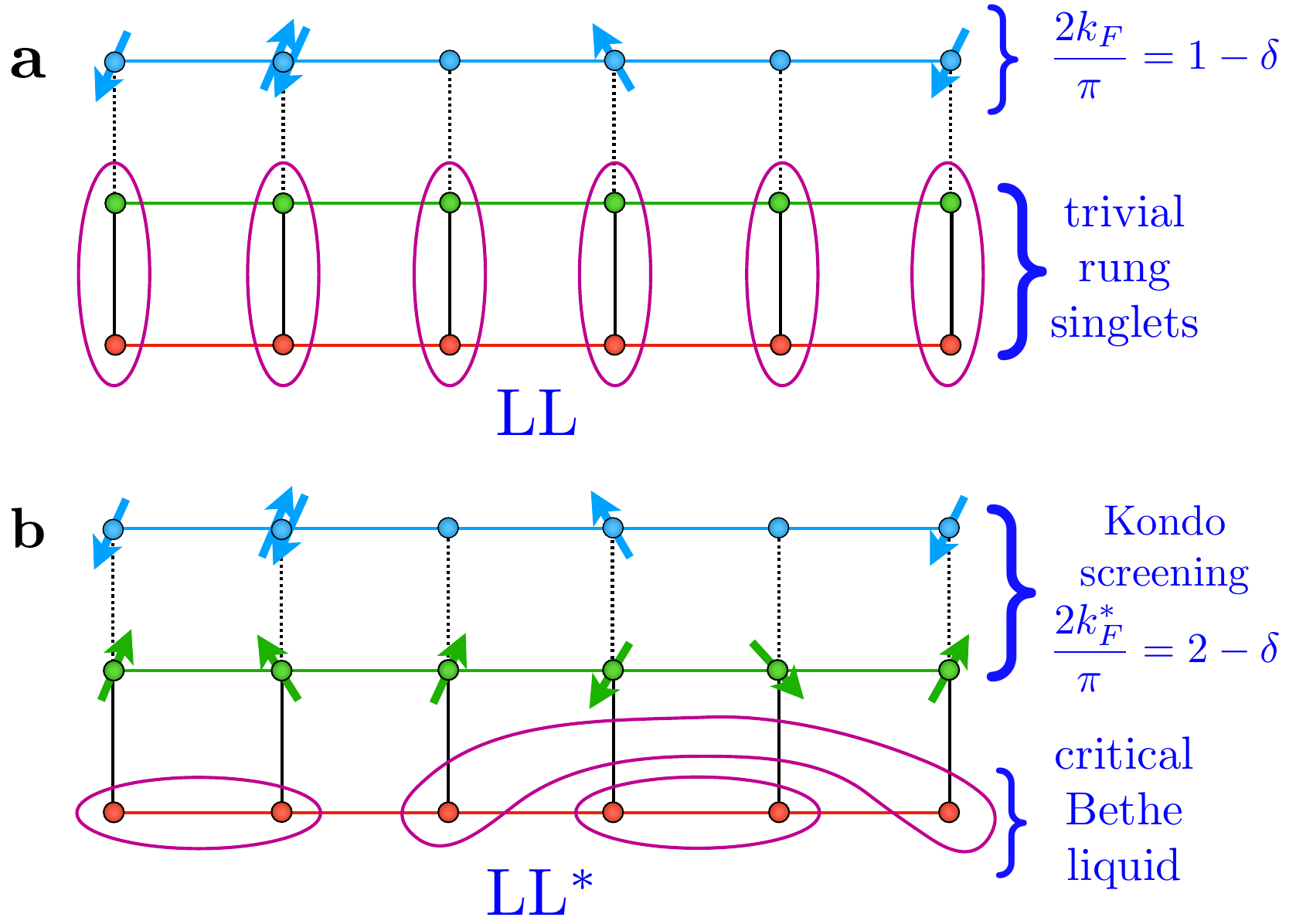}}
        \caption{Schematic representation of two phases of the one-dimensional ALM. \textbf{Panel a:} The mobile fermions form a Luttinger liquid with the conventional Fermi wavevector, $k_F$, of a decoupled chain, while the spin ladder is smoothly connected to the trivial, gapped rung-singlet phase. \textbf{Panel b:} The mobile fermions are Kondo screened, and the Fermi wavevector, $k_F^\ast$, is that of a Kondo lattice model; the bottom leg of the spin ladder forms a critical spin liquid with the same critical singularities as those of the decoupled spin-$1/2$ chain solved by Bethe. 
        \label{fig:alm_schematic} }
    \end{center}
\end{figure}

\begin{figure}[t]
    \begin{center}
\centerline{\includegraphics[width=\columnwidth]{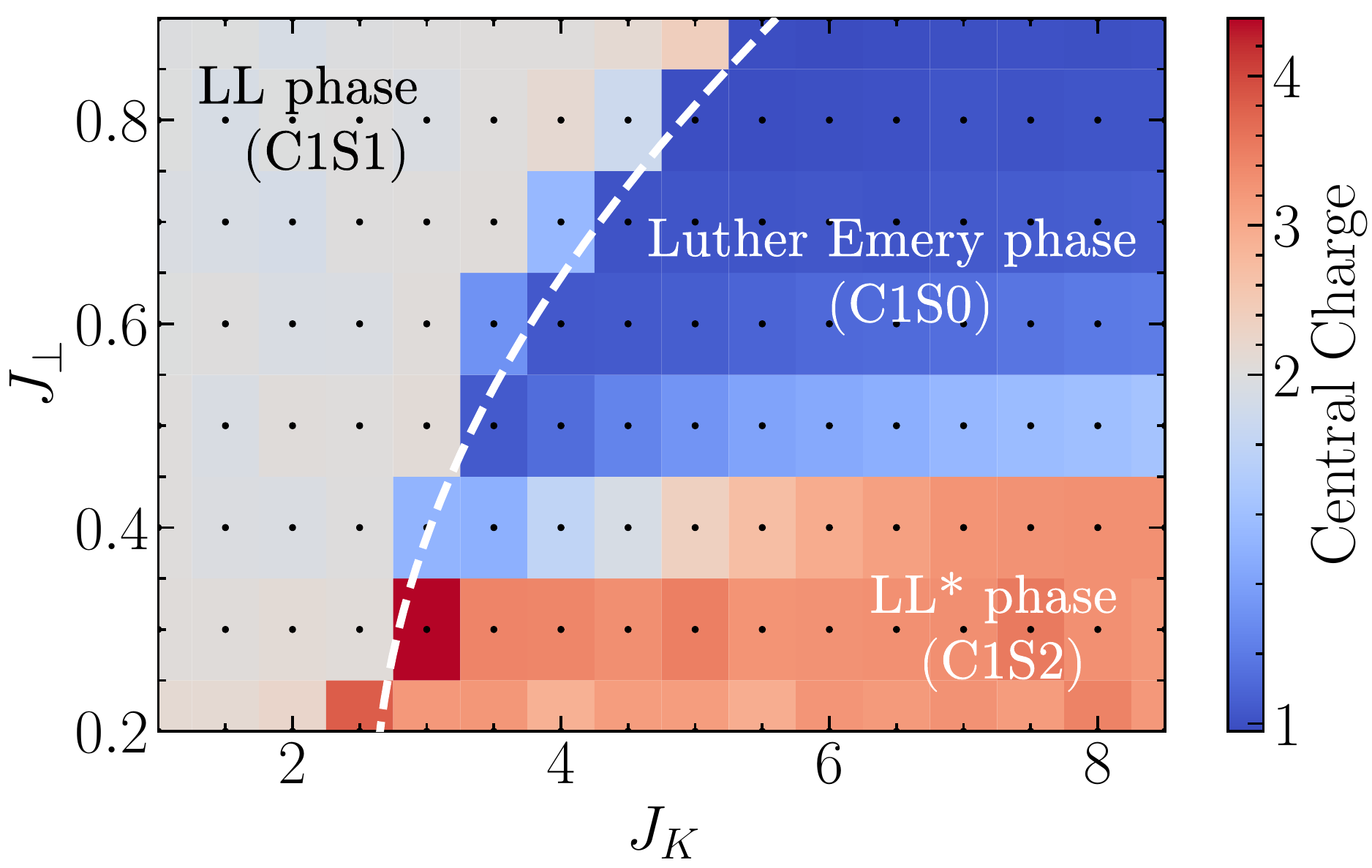}}
        \caption{\label{fig:phase_diagram} Phase diagram of the one-dimensional ALM as a function of the Kondo coupling $J_K$ and the interlayer exchange $J_\perp$ obtained from iDMRG calculations.
        The gray region is the LL phase (C1S1) with central charge $c=2$, the red region is the LL$^*$ phase (C1S2) with $c=3$, while the blue region is the Luther-Emery phase with $c=1$ (C1S0). The dashed white line denotes the boundary of the Kondo screened phase with peak at $2k_F^*$ in spin-spin correlation functions (refer to \cref{sec:energy_corr}).
         The central charge is extracted from the entanglement entropy ($\chi=500,1000$) with a unit cell of length $L=14$ with number of electrons $N_e=10$ ($\delta \approx 0.2857$) for fixed values of $t=1.0$ and $J_1=J_2=0.5$.}
    \end{center}
\end{figure}

Two of the expected ground states of the one-dimensional ALM are sketched in Fig.~\ref{fig:alm_schematic}. The LL phase is smoothly connected to the regime where the chain and spin-ladder are decoupled, and the chain forms a Luttinger liquid with the Fermi wavevector $k_F$. In the LL* phase, the Fermi wavevector $k_F^\ast$ shifts to that of the Kondo-screened phase of a Kondo lattice, while the bottom leg of the spin ladder forms a critical Bethe spin liquid. From our studies of the central charge of the ground state in the Density Matrix Renormalization Group (DMRG) computation, we find evidence for the instability of a significant portion of the LL* phase to a Luther-Emery (LE) phase \cite{Voit95}, as shown in the phase diagram in \cref{fig:phase_diagram}.

The two-dimensional version of the ALM was extensively studied in the context of high-$T_c$ superconductors~\cite{Zhang2020,Zhang2020_2,Maria21,Mascot2022,Nikolaenko2023,Christos2023,Christos2024,BCS24,Pandey2025,Boulder25,Coleman2025,Coleman2026}. It was demonstrated that the model reproduces many salient features of the cuprate phase diagram, such as the volume breaking transition, the emergence of the Fermi arcs, and non-Fermi liquid behavior. The LL* phase in one dimension is the analog of the fractionalized Fermi liquid (FL*) phase in two dimensions, which has been proposed as a description of the pseudogap regime in the cuprates and can emerge in theories of Kondo breakdown. In this context, it is interesting that we find an instability of LL* to the LE liquid in one dimension, which could be analogous to the onset of superconductivity from the pseudogap phase of the cuprates.
Most of the approaches to the ALM in two dimensions involved parton-decomposition, applied originally in the Kondo model~\cite{Read1983,Coleman1987}, while the exact treatment of the model was never performed. 

In one dimension, related numerical studies of the Kondo-Heisenberg model~\cite{Nikolaenko2024}, which consists of a fermionic
chain coupled to a single spin layer, based on DMRG \cite{white1992}, have discovered two distinct metallic phases with a jump in the $2k_F$ when tuning the Kondo coupling. We expect an even richer phase diagram to emerge for the ALM as the Kondo coupling $J_K$ and the inter-layer spin coupling $J_\perp$ are varied. 
In this context, the ALM on a chain [see \cref{eq:hamiltonian}] provides a controlled setting in which itinerant fermions interact with multiple localized spin degrees of freedom, giving rise to a rich interplay between screening mechanisms and enabling the study of the transition from underscreened to overscreened regimes. At the same time, the availability of high-precision DMRG results makes this system a well-defined benchmark for assessing the accuracy of NQS-based variational approaches.
Establishing the accuracy of the approach in one dimension, where DMRG benchmarks are available across multiple phases, is a necessary step before extending the method to the more challenging two-dimensional case.


A direct comparison with DMRG under open boundary conditions shows that the Transformer-based variational wavefunction introduced here achieves excellent quantitative agreement for both ground-state energies and correlation functions. This accuracy enables a reliable characterization of the ground-state phase diagram, which has not been systematically studied so far.

Crucially, the NQS Ansatz treats open and periodic boundary conditions on the same footing while maintaining comparable accuracy. By contrast, obtaining highly accurate DMRG results for large systems with periodic boundary conditions is computationally demanding~\cite{verstraete2004, sandvik2007, pippan2010}. The main advantages of PBC are a substantial reduction of finite-size effects in the scaling of physical observables and direct access to experimentally relevant quantities, such as the lowest-energy triplet excitation at finite momentum~\cite{choo2018, nomura2021, viteritti2022, viteritti2025prb}. More importantly, this technique can also be extended to clusters in higher dimensions, without the theoretical and algorithmic constraints that often limit tensor-network methods in fully two-dimensional settings~\cite{Schollwock2011, stoudenmire2012, orus2014}. Exploiting this, we take a first step beyond one dimension and benchmark the method directly on an $8\times8$ square lattice, showing that the same architecture reproduces the two analytically understood limits of the model in quantitative agreement with quantum Monte Carlo~\cite{qmcheisenbergbenchmark}.

The remainder of the paper is organized as follows. In \cref{sec:methods}, we present the Transformer-based variational wave function and discuss its architecture and implementation. \cref{sec:results} contains the main results, including benchmarks against DMRG, an analysis of correlation functions and the excitation spectrum, the characterization of the phase diagram, and a discussion of the different physical regimes. In \cref{sec:benchmark2d}, we benchmark the method in two dimensions, verifying that it reproduces the small- and large-Fermi surface limits of the model. Finally, \cref{sec:conclusions} summarizes our findings and outlines perspectives for extending this approach to higher-dimensional systems and more challenging lattice models.

\section{Methods}\label{sec:methods}
In this Section, we introduce a variational wavefunction designed for lattice systems whose local Hilbert space includes both spin and fermionic degrees of freedom. Neural-Network Quantum States provide compact representations of quantum many-body wavefunctions by employing a neural network with parameters $\theta$ to map input basis states $\boldsymbol{s}$ to real-valued amplitudes $\Psi_\theta(\boldsymbol{s})$~\cite{carleo2017, lange2024}. For a lattice model with composite degrees of freedom at each site, an input basis state is specified as the sequence $\boldsymbol{s} = (s_1, \ldots, s_N)$, where $s_i$ denotes the local configuration at site $i$ and $N$ is the total number of lattice sites.

In the Ancilla Layer model introduced in \cref{eq:hamiltonian}, the local configuration at each site is given by the tuple $s_i = (n_{i\uparrow}, n_{i\downarrow}, S^z_{1i}, S^z_{2i})$, where $n_{i\sigma} \in \{0,1\}$ with $\sigma \in \{\uparrow, \downarrow\}$ denotes the occupation number of spinful fermions, and $S^z_{1i}, S^z_{2i} = \pm \tfrac{1}{2}$ are the $z$-components of two ancillary spins. The resulting local Hilbert space has dimension $\mathcal{V} = 2^4$ (see \cref{fig:alm_model}(b) for a pictorial representation).

By assigning a unique integer label to each local configuration, each basis state can be equivalently represented as a sequence of integers $\boldsymbol{t}=(t_1, \ldots, t_N)$, with ${t_i \in \{0, 1, \ldots, \mathcal{V}-1\}}$. Here, we adopt the notation $\boldsymbol{t}$ by analogy with natural language processing, where sentences composed of discrete tokens are represented as sequences of integers through a process known as \emph{tokenization}~\cite{jm3,vaswani2017attention}. The Transformer architecture, originally introduced in Ref.~\cite{vaswani2017attention}, is explicitly designed to process such kinds of sequences.

Each token $t_i$ is first mapped to a vector $\boldsymbol{x}_i \in \mathbb{R}^d$ via an embedding lookup table implemented as a trainable matrix of shape $\mathcal{V} \times d$, where $d$ is the embedding dimension, a hyperparameter of the network (refer to \cref{sec:architecture} of the \textit{Appendix} for additional details). These embedding vectors can be interpreted as abstract representations of the local Hilbert-space configurations at each site. The sequence $(\boldsymbol{x}_1, \ldots, \boldsymbol{x}_N)$ constitutes the input to the Transformer. Notice that at this stage, each vector depends \textit{only} on the corresponding local configuration, namely $\boldsymbol{x}_i = \boldsymbol{x}_i(s_i)$.

The central role of the Transformer architecture, and in particular its self-attention mechanism, is to map the input sequence to a new sequence in which each element incorporates information from the entire input. Specifically, the self-attention mechanism produces output vectors $\boldsymbol{A}_i = \boldsymbol{A}_i(\boldsymbol{s}) \in \mathbb{R}^d$ with $i=1,\dots, N$, which are commonly referred to as \emph{context-aware} representations of the $i$-th input token. In natural language processing, such representations reflect the fact that the meaning of a word depends on its surrounding context; analogously, in this setting, they encode non-local correlations between degrees of freedom at different lattice sites.

\begin{figure}[t]
    \begin{center}
\centerline{\includegraphics[width=0.8\columnwidth]{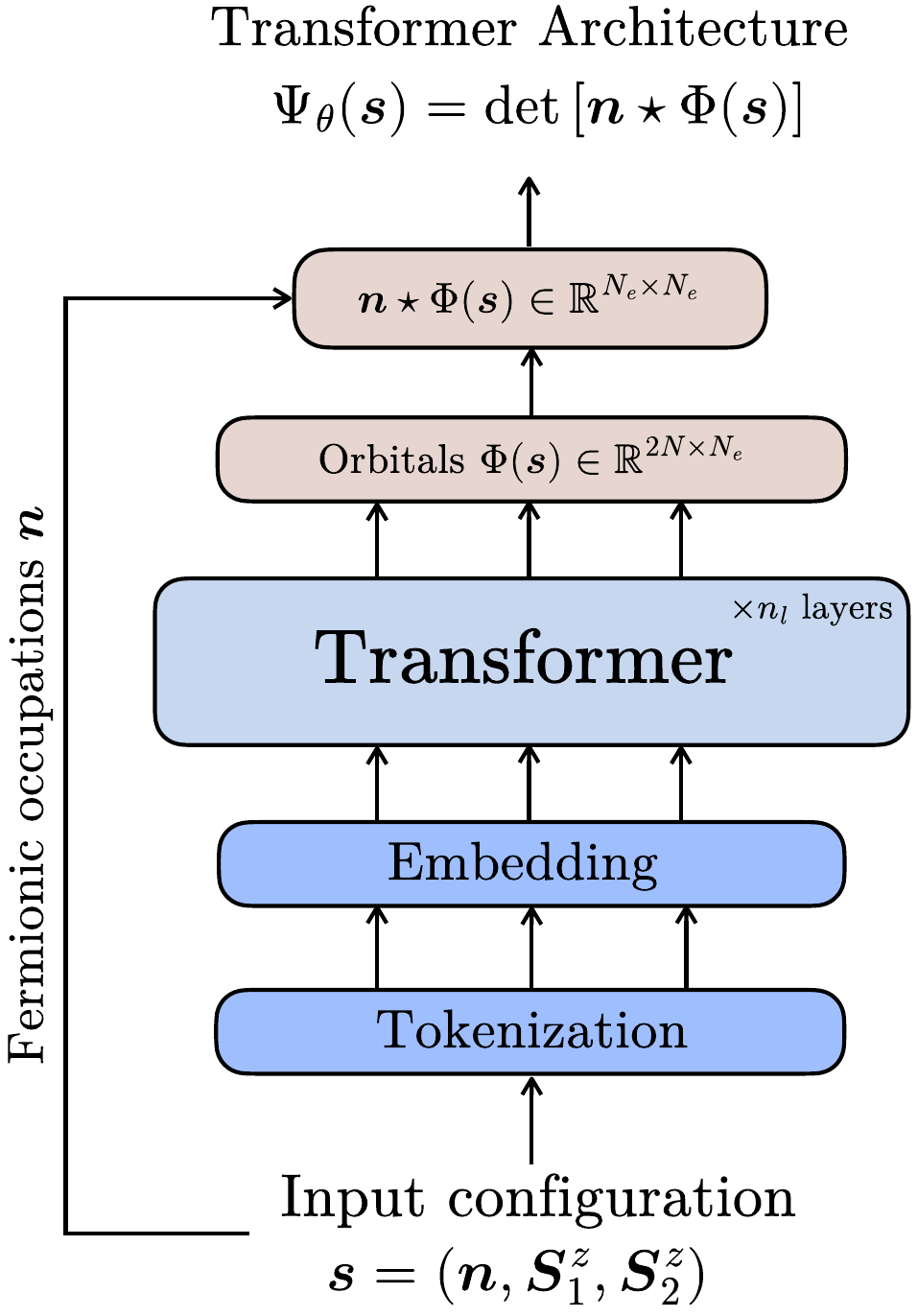}}
        \caption{\label{fig:architecture}A physical configuration $\boldsymbol{s}=(s_1,\ldots,s_N)$, with
$s_i=(n_{i\uparrow},n_{i\downarrow},S^z_{1i},S^z_{2i})$, is first encoded as a
sequence of integer tokens $(t_1,\ldots,t_N)$, where $t_i\in\{0,\ldots,\mathcal{V}-1\}$
and $\mathcal{V}$ is the local Hilbert-space dimension. The tokens are embedded
into feature vectors $(\boldsymbol{x}_1,\ldots,\boldsymbol{x}_N)$ and processed by
a Transformer to produce context-dependent outputs $(\boldsymbol{y}_1,\ldots,\boldsymbol{y}_N)$.
These outputs parameterize the backflow single-particle orbitals $\Phi(\boldsymbol{s})$,
and the many-body amplitude $\Psi_\theta(\boldsymbol{s})$ is obtained
via a Slater-determinant construction.}
    \end{center}
\end{figure}

In the standard self-attention mechanism introduced by~\citet{vaswani2017attention}, the output vectors are computed according to the rule ${\boldsymbol{A}_i = \sum_{j=1}^N \alpha_{ij}(\boldsymbol{x}_i, \boldsymbol{x}_j)\, V \boldsymbol{x}_j}$, where $V \in \mathbb{R}^{d \times d}$ is a trainable linear transformation applied independently to each input vector, and the matrix $\alpha_{ij}(\boldsymbol{x}_i, \boldsymbol{x}_j) \in \mathbb{R}^{N\times N}$ denotes the attention weights. These weights quantify the relevance of site $j$ to site $i$, and in the original formulation, are learned functions of the corresponding input embeddings. In this work, we instead adopt a simplified variant known as the \emph{Factored attention mechanism}~\cite{jelassi2022,  rende2024prr, viteritti2023prl, viteritti2025prb}, in which the attention weights $\alpha_{ij}$ depend only on the site indices and are parameterized as a trainable $N \times N$ matrix. As shown in Ref.~\cite{rende2025iop}, for spin systems this factorization significantly reduces the computational cost of the attention mechanism without compromising the accuracy of the resulting variational state. Furthermore, we incorporate a spatial bias, following Ref.~\cite{viteritti2026}, which enforces a decay of the attention weights with increasing distance between sites $i$ and $j$, thereby encoding information about the underlying lattice structure directly into the architecture.

In a deep Transformer architecture, the self-attention mechanism is applied repeatedly across multiple layers. In addition, several architectural components are employed to enhance expressivity and stabilize training, including Multi-Head attention~\cite{vaswani2017attention}, fully connected networks, pre-layer normalization~\cite{prelayernorm}, and residual skip connections~\cite{resnet}. A detailed description is provided in Ref.~\cite{viteritti2025prb}.

\begin{figure}[t]
    \begin{center}
\centerline{\includegraphics[width=\columnwidth]{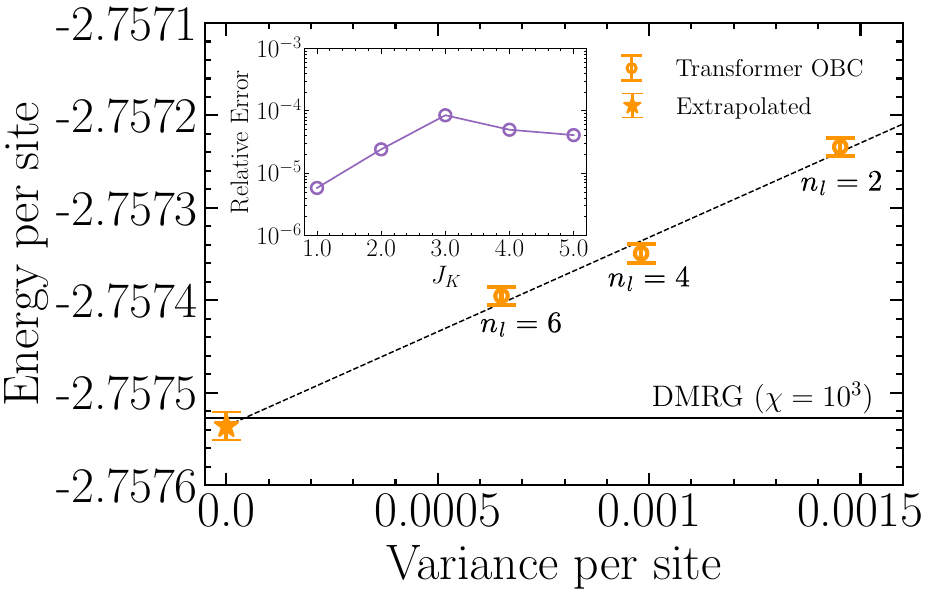}}
        \caption{\label{fig:energy_comparison_DMRG} Variational energy per site as a function of the energy variance per site on a chain of $N=42$ sites with OBC. The couplings are $t=1.0$, $J_1=J_2=J_\perp=0.5$, $J_K=4.0$ and the number of electrons is $N_e=30$ ($\delta \approx 0.2857$). Results are shown for Transformer wave functions with different number of layers $n_l=2,4,6$ (refer to \cref{sec:architecture} for details). The dashed line indicates the DMRG reference energy obtained with bond dimension $\chi=10^3$. \textbf{Inset}: Relative energy error $\Delta\varepsilon = \tfrac{|E_{\text{NQS}} - E_{\text{DMRG}}|}{|E_{\text{DMRG}}|}$ as a function of the Kondo coupling $J_K$, considering a Transformer with $n_l=4$ layers. }
    \end{center}
\end{figure}

The output of the Transformer architecture after the application of multiple layers is a sequence of vectors $(\boldsymbol{y}_1, \ldots, \boldsymbol{y}_N)$, with $\boldsymbol{y}_i \in \mathbb{R}^d$, which we use to construct a set of effective single-particle orbitals defining a backflow transformation for the fermionic degrees of freedom~\cite{diluo_prl}. The backflow orbitals are obtained through a site-resolved linear mapping of the Transformer outputs, yielding the tensor ${\Phi_{i\sigma\alpha} = \sum_{\beta=1}^d y_{i\beta} W_{i\sigma\alpha\beta}}$, where $W_{i\sigma\alpha\beta}$ are trainable parameters and $\alpha = 1, \ldots, N_e$ labels the single-particle orbitals, with $N_e$ representing the total number of fermions in the system~\cite{gu2025solvinghubbardmodelneural, ma2025}. To construct the final backflow matrix, we introduce the composite index $r = (i,\sigma)$, which combines the lattice site $i$ and the fermionic spin degree of freedom $\sigma$, and reshape $\Phi_{i\sigma\alpha}$ into a matrix $\Phi_{r\alpha} \in \mathbb{R}^{2N \times N_e}$. At the end, the many-body wavefunction amplitude is obtained by selecting the rows of this matrix corresponding to the occupied fermionic sites and computing the determinant of the resulting matrix. Namely, the final amplitude is given by ${\Psi_{\theta}(\boldsymbol{s}) = \det \left[ \boldsymbol{n} \star \Phi(\boldsymbol{s} )\right]}$, where $\boldsymbol{n} \star \Phi$ denotes the $N_e \times N_e$ submatrix of $\Phi$ indexed by the occupation numbers of $\boldsymbol{n}=(n_{1\uparrow},\dots, n_{N\uparrow},n_{1\downarrow},\dots, n_{N\downarrow})$. For a schematic illustration of the full construction of the wave function refer to \cref{fig:architecture}. The hyperparameters of the architecture and the optimization details are provided in the \cref{sec:architecture} of the \textit{Appendix}. 

We emphasize that this construction differs qualitatively from previous variational treatments of systems with coexisting spin and fermionic degrees of freedom, where the wave function is typically written as a product of a fermionic amplitude and a separate spin factor~\cite{nomura2020, marino2025, piccioni2025, favata2025, mahajan2025}, so that the two sectors are factorized by construction. Here, the methodological key innovation is that no such factorization is imposed: the Transformer processes the full composite configuration and predicts backflow orbitals $\Phi(\boldsymbol{s})$ that depend explicitly on the spin degrees of freedom, so that the resulting state naturally couples the fermionic and spin sectors. Related to this choice is the tokenization scheme, which combines all local degrees of freedom on each site, the fermionic occupation and the two ancilla spins, into a single token. This is not the only possibility: one could instead assign a separate token to each degree of freedom, at the cost of a longer input sequence ($3N$ instead of $N$). Since the computational cost of the Transformer scales linearly with the sequence length and is essentially independent of the number of states per token (see Ref.~\cite{rende2026scalinglawsneuralnetworkquantum}), this alternative is about three times more expensive; we verified that, despite this higher cost, it is no more accurate than the single-token encoding, and in fact converges more slowly in the strongly interacting regime at intermediate $J_K$.

\section{Results}\label{sec:results}
In the following, unless stated otherwise, we consider a system of ${N = 42}$ sites with $N_e = 30$ fermions, corresponding to a hole doping of $\delta \approx 0.2857$.

\subsection{Energy and Correlation Functions}\label{sec:energy_corr}
We begin by benchmarking the ALM on a cluster under open boundary conditions, comparing the results of the Transformer-based wavefunction against DMRG results. We set $J_1 = J_2 = J_\perp = 0.5$ and $J_K = 4.0$. First, we assess the accuracy of our method by calculating the variational energy as a function of the variance, increasing the number of parameters of the neural network by varying the number of Transformer layers $n_l = 2, 4, 6$ (see \cref{fig:energy_comparison_DMRG}). Increasing the depth of the network systematically improves the variational accuracy. Extrapolating the energy to zero variance~\cite{becca2015} yields values in excellent agreement with the DMRG reference, computed with bond dimension $\chi = 10^3$. In the inset of \cref{fig:energy_comparison_DMRG}, we further assess the accuracy of the Transformer wavefunction as a function of the Kondo coupling $J_K$, reporting the relative energy error with respect to DMRG, defined as $\Delta\varepsilon = \tfrac{|E_{\text{NQS}} - E_{\text{DMRG}}|}{|E_{\text{DMRG}}|}$. Across the entire range of $J_K$ considered, the relative error remains below $\Delta\varepsilon \le 10^{-4}$, with a modest maximum around $J_K \approx 3.0$, demonstrating the robustness of the approach across different coupling regimes.

\begin{figure}[t]
    \begin{center}
\centerline{\includegraphics[width=\columnwidth]{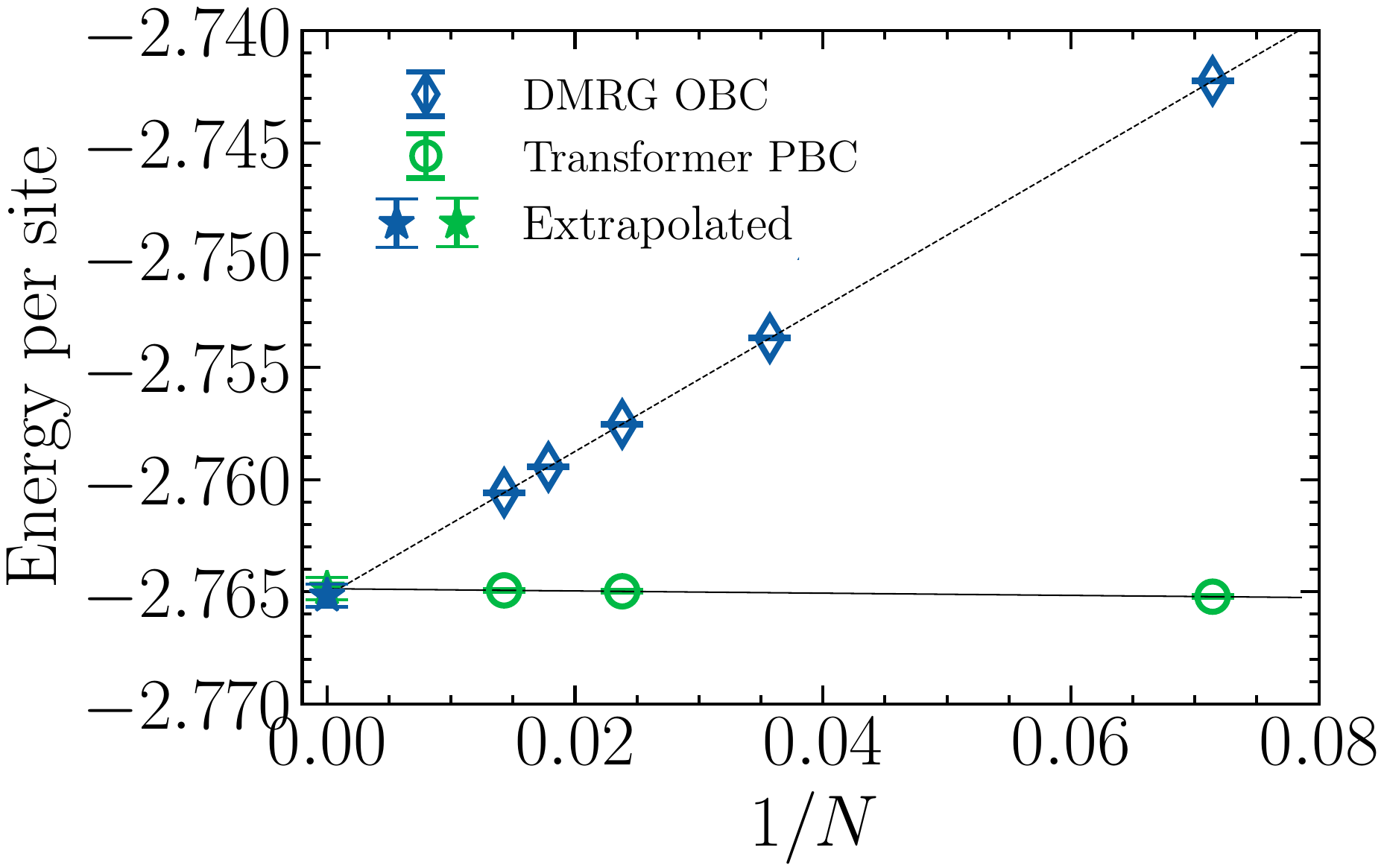}}
        \caption{\label{fig:scalings_obc_pbc} Finite-size scaling of the ground-state energy per site from $N=14$ to $N=70$ sites fixing the doping at ${\delta\approx 0.2857}$ as a function of inverse system size $1/N$. The couplings are $t=1.0$, $J_1=J_2=J_\perp=0.5$ and $J_K=4.0$. Results obtained with DMRG for open boundary conditions (OBC) are compared with Transformer variational energies computed with periodic boundary conditions (PBC).}
    \end{center}
\end{figure}
\begin{figure*}[ht]
    \begin{center}
\centerline{\includegraphics[width=2\columnwidth]{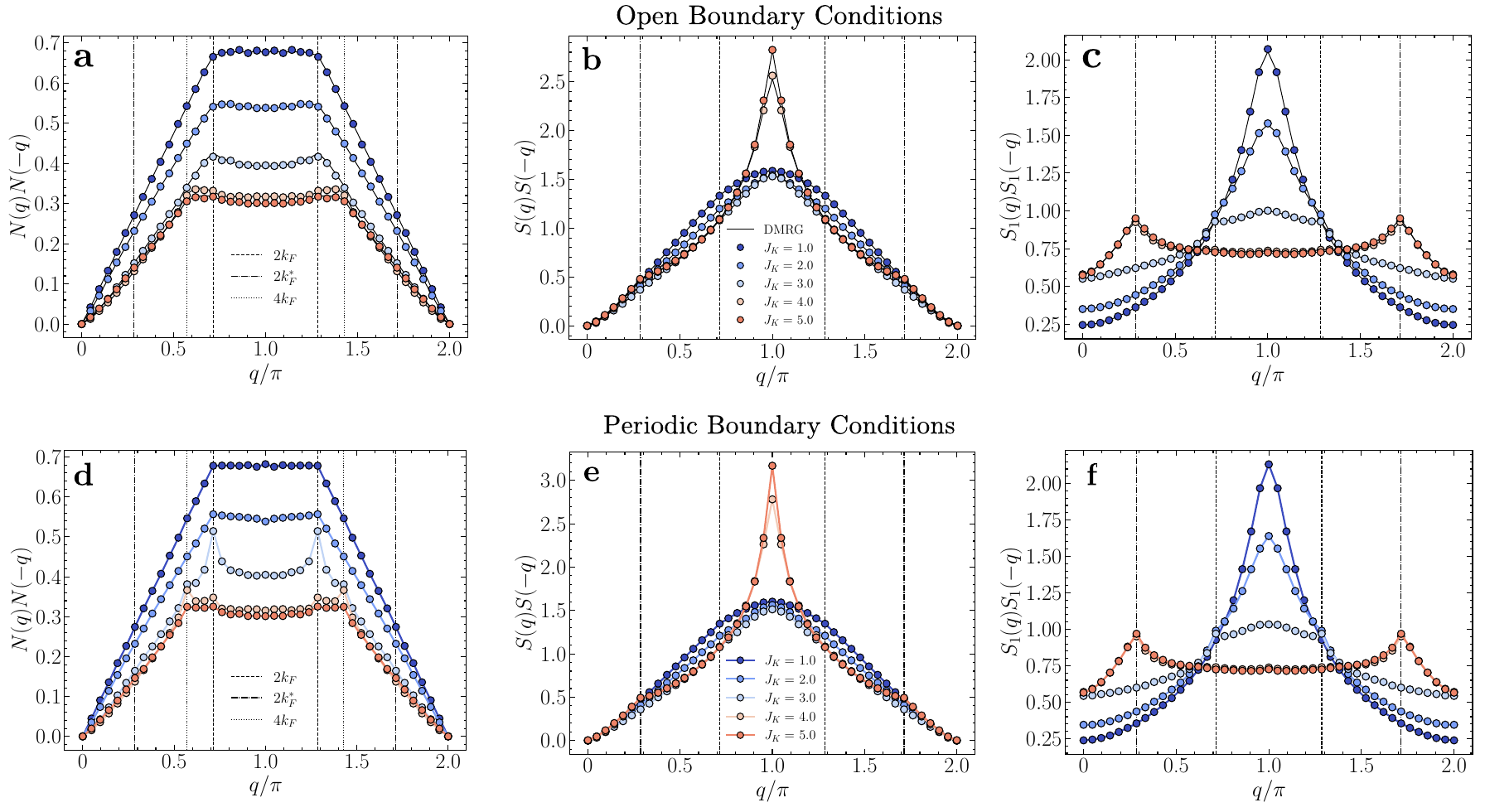}}
        \caption{\label{fig:correlations} Momentum-resolved correlation functions at hole doping $\delta \approx 0.2857$
        for a chain of length $N=42$, shown for several values of the Kondo coupling $J_K$ at fixed $t=1$, $J_1=J_2=J_\perp=0.5$. The three columns report, from left to right,
        the charge structure factor $N(q)N(-q)$, the total spin structure factor $S(q)S(-q)$,
        and the spin structure factor of the first ancilla chain $S_1(q)S_1(-q)$.
        \textbf{Top panels a, b, c:} Circles denote results obtained in open boundary conditions from the Transformer-based variational wave function, while solid lines show the
        corresponding DMRG reference data. \textbf{Bottom panels d, e, f:} The structure factors are computed using the Transformer wave function in periodic boundary conditions.
        In all panels, vertical guide lines mark the characteristic wavevectors $2k_F$
        (dashed), $2k_F^{\ast}$ (dash-dotted), and $4k_F$ (dotted).}
    \end{center}
\end{figure*}

We next examine the finite-size scaling of the ground-state energy under open and periodic boundary conditions. \cref{fig:scalings_obc_pbc} displays the energy per site from $N=14$ to $N=70$ as a function of $1/N$ at fixed coupling $J_K = 4.0$ and doping $\delta \approx 0.2857$. For open boundaries, the DMRG data exhibit a clear linear dependence on $1/N$, enabling a controlled extrapolation to the thermodynamic limit. By contrast, the Transformer results with periodic boundaries show a much weaker size dependence, with energies already close to their asymptotic value for the largest systems considered. Despite these different scaling behaviors, extrapolation yields a consistent thermodynamic-limit energy within numerical accuracy for both boundary conditions. This behavior highlights a key advantage of the Transformer-based variational approach. While OBC introduce sizable edge effects that lead to algebraic finite-size corrections in the energy, PBC strongly suppress such effects, resulting in faster convergence with system size.

We now turn to the analysis of correlation functions. Given a local operator $\hat{A}_i$ acting on site $i$, we define the correlation matrix $C_{ij}=\braket{\hat{A}_i \hat{A}_j} - \braket{\hat{A}_i}\braket{ \hat{A}_j}$ with ${i,j=0,\dots,N-1}$, where the expectation values are evaluated on the variational state. The corresponding structure factor is defined as
\begin{equation}\label{eq:fourier_transform}
    C(q)C(-q)=\frac{1}{N}\sum_{i,j=0}^{N-1} e^{\, i q (i-j)}\, C_{ij}\, ,
\end{equation}
with $q=\tfrac{2\pi}{N}n$ and $n=0,1,\dots,N-1$, the allowed momenta on a chain of $N$ sites. This definition can be used for both open and periodic boundary conditions.

In the following, we focus on three choices of $\hat{A}_i$. We consider the local charge operator $\hat{N}_i=\sum_{\sigma}\hat{c}^\dagger_{i\sigma}\hat{c}_{i\sigma}$, the spin operator restricted to the first ancilla chain $\hat{\boldsymbol{S}}_{i,1}$, and the total on-site spin operator
\begin{equation}
    \hat{\boldsymbol{S}}_i
    =
    \frac{1}{2} \sum_{\sigma,\sigma'=\uparrow,\downarrow}
    \hat{c}^\dagger_{i\sigma}\,\boldsymbol{\tau}_{\sigma\sigma'}\,\hat{c}_{i\sigma'}
    +
    \hat{\boldsymbol{S}}_{i,1}
    +
    \hat{\boldsymbol{S}}_{i,2}\, ,
\end{equation}
where $\boldsymbol{\tau}$ denotes the vector of Pauli matrices. We analyze the corresponding charge structure factor $N(q)N(-q)$, the spin structure factor of the first ancilla chain $S_1(q)S_1(-q)$, and the total spin structure factor $S(q)S(-q)$, all defined via \cref{eq:fourier_transform}.

In the top panels of \cref{fig:correlations}, we show the plane-wave–projected structure factor for OBC, benchmarking the Transformer results with DMRG. 

For small Kondo coupling ($J_K=1.0, 2.0$), the electronic and ancillary spin degrees of freedom are only weakly correlated. In this regime, the first layer is well approximated by free electrons, forming a Luttinger liquid  with Fermi momentum $k_F$, while the ancillary spins form rung-singlets with a finite spin gap and effectively decouple from the electrons~\cite{Barnes1993}. This behavior is reflected in panel (a), in which the charge structure factor $N(q)N(-q)$ exhibits the characteristic free-electron form with a cusp at 
\begin{equation}
\frac{2k_F}{\pi} = 1 - \delta\,.     
\end{equation}
In panel (b), the total spin structure factor $S(q)S(-q)$ is peaked at $q=\pi$, reflecting dominant antiferromagnetic correlations. Consistently, in panel (c), the structure factor of the first ancilla, $S_1(q)S_1(-q)$, is also peaked at $q=\pi$, indicating the absence of Kondo screening. At small momenta, it vanishes as $q^2$, in agreement with the presence of a spin gap in the ancillary sector. In this weak-coupling phase, the Transformer results are in excellent quantitative agreement with DMRG for all observables considered.

At larger values of the Kondo coupling ($J_K=4.0, 5.0$), the system enters a Kondo-screened regime in which the electronic spins effectively hybridize with and screen the spins of the first ancilla layer. As a result of hybridization, a Fermi surface with a Fermi momentum $k_F^{\ast}$ is formed, incorporating the spins from the first layer in the Fermi surface volume. As visible in Fig.~\ref{fig:correlations}, this regime is characterized by a redistribution of spectral weight in the spin sector. In panel (b), the total spin structure factor $S(q)S(-q)$ shows that the pronounced peak at $q=\pi$ is strongly suppressed, indicating that antiferromagnetic fluctuations are no longer dominant. Furthermore, in panel (c), the $S_1(q)S_1(-q)$ spin structure factor develops a sharp peak at  
\begin{equation}
    \frac{2 k_F^{\ast}}{\pi}=2-\delta\,,
\end{equation} 
which confirms the establishment of the Kondo screened phase. Again, the Transformer wave function accurately captures these features, reproducing both the momentum dependence and relative magnitude of the correlation functions in the different regimes.

Close to the critical point ($J_K=3$), the spin structure factor $S_1(q)S_1(-q)$ demonstrates neither a strong peak at $q=\pi$ nor an increased response at $q=2k_F^{\ast}$. In the vicinity of the transition, most correlation lengths become extremely short, on the order of a few lattice spacings. Furthermore, the critical point might be surrounded by another phase. Similar behavior was observed in the Kondo-Heisenberg model~\cite{Nikolaenko2024}, and it is believed to be related to local criticality. We discuss in more detail the nature of the transition in \cref{app:transition} of the \textit{Appendix}.

The bottom panels of \cref{fig:correlations}, namely panels (d), (e) and (f), show the same correlation functions with periodic boundary conditions, computed only with the Transformer Ansatz, since DMRG becomes substantially more demanding in this setup. Note that near the phase transition ($J_K=3.0$), density-density correlation functions exhibit stronger peaks at $2k_F$ and  $4k_F$ [see panel (d)]. This can be explained in the following way.
For periodic boundary conditions, translational invariance ensures that momentum remains a good quantum number, concentrating the $2k_F$ and $4k_F$ spectral weight at well-defined momenta. In contrast, open boundaries break translational symmetry and redistribute this weight among neighboring momenta through boundary-induced mixing, resulting in broader and generally less pronounced peaks in the connected correlator. Interestingly, for the disconnected charge correlator, similar strong peaks emerge due to Friedel oscillations of the density profile $\langle \hat{N}_i \rangle$.

The spin-spin correlations are similar in shape and amplitude compared to open boundary conditions. Importantly, switching from open to periodic boundary conditions does not modify the accuracy of the NQS state, which continues to provide quantitatively reliable correlation functions in the PBC geometry.

\subsection{Phase Diagram}
In \cref{fig:phase_diagram} of Section~\ref{sec:intro} we presented the phase diagram of the one-dimensional ALM as a function of the Kondo coupling $J_K$ and the interlayer exchange $J_\perp$, obtained from the values of the central charge. The central charge $c$ is extracted via the infinite-size variant of the DMRG (iDMRG) algorithm~\cite{mcculloch2008} using the entanglement entropy of a half-infinite chain $S=\tfrac{c}{6} \ln[\xi(\chi)]$, where $\xi(\chi)$ is the largest correlation length at a  fixed bond dimension $\chi$~\cite{Pollmann2009,Calabrese2004}.

As discussed in \cref{sec:energy_corr}, for small $J_K$, the first layer effectively decouples from the rest of the system, while the ancillary spins form rung singlets and exhibit strong antiferromagnetic fluctuations at $q=\pi$ [see Fig.~\ref{fig:alm_schematic}(a)]. In this regime, the system realizes a Luttinger liquid with Fermi momentum $k_F$ and one charge and one spin mode (C1S1) and central charge $c=2$. These modes come from the first layer only since the ancillary spin degrees of freedom form spin gapped rung-singlets on each site and decouple from the electrons~\cite{Barnes1993}.

At large $J_K$, the system enters a Kondo-screened phase in which the spins of the first ancillary layer hybridize with the conduction electrons. 
At small $J_\perp$, the second ancillary layer enables the fractionalization of the gapless mode. Specifically, the mode splits into a $2k_F^{\ast}$ contribution from the hybridized layers and a $q=\pi$ contribution from the second ancillary layer, forming a phase we call LL$^*$~\cite{Zhang2021} [see Fig.~\ref{fig:alm_schematic}(b)]. The phase has one charge and two spin modes (one arising from hybridized levels and the other from the second ancillary layer), which is denoted as C1S2 and has central charge $c=3$.
Note that the total momentum of the gapless mode is in accordance with the Lieb-Schultz-Mattis theorem~\cite{Lieb1961,Yamanaka1997}. In two dimensions, the LL$^*$ phase descends into an FL$^*$ (Fractionalized Fermi liquid~\cite{Senthil2003}), where the conventional Luttinger volume constraint is not satisfied. That is because the second ancillary layer hosts a spin liquid alongside the large Fermi surface formed by the other layers.

For larger $J_\perp$, the LL$^* $ phase is generally unstable toward a Luther-Emery phase~\cite{Nikolaenko2024,Berg2010,Zhang2022}, see \cref{app:transition} of the \textit{Appendix} for details. Accordingly, two spin modes become gapped, with only one charge mode left; the phase is C1S0 with the central charge $c=1$.
We note that increasing $J_\perp$ increases the spin gap in the ancillary spin layers, as the triplet gap is on the order of $J_\perp$. Therefore, the critical point moves to larger values of $J_K$ since it becomes harder for electrons to hybridize with the first ancillary spin layer. 

Alternatively, the phase boundaries can be determined from the spin-spin correlation functions. 
 In particular, in the Kondo-screened phase $S_1(q)S_1(-q)$ develops a pronounced peak at Fermi momentum $q/\pi=2k_F^{\ast}/\pi=2-\delta$ (see \cref{fig:correlations}), signaling the enlargement of the Fermi surface. The dashed white line in \cref{fig:phase_diagram} shows the phase boundary identified by analyzing the relative magnitude of the second derivative of the spin structure factor $S_1(q)S_1(-q)$ at momenta $q=\pi$ and $q=2k_F^{\ast}$. This results in a phase boundary that matches the one determined from the central charge. At the triple critical point where three phases converge, we observe a central charge $c > 4$ that persists even at a bond dimension of $\chi = 2000$. This behavior may signal a novel critical point, aligning with observations previously reported in~\cite{Nikolaenko2024}.

\begin{figure}[t]
    \begin{center}
\centerline{\includegraphics[width=\columnwidth]{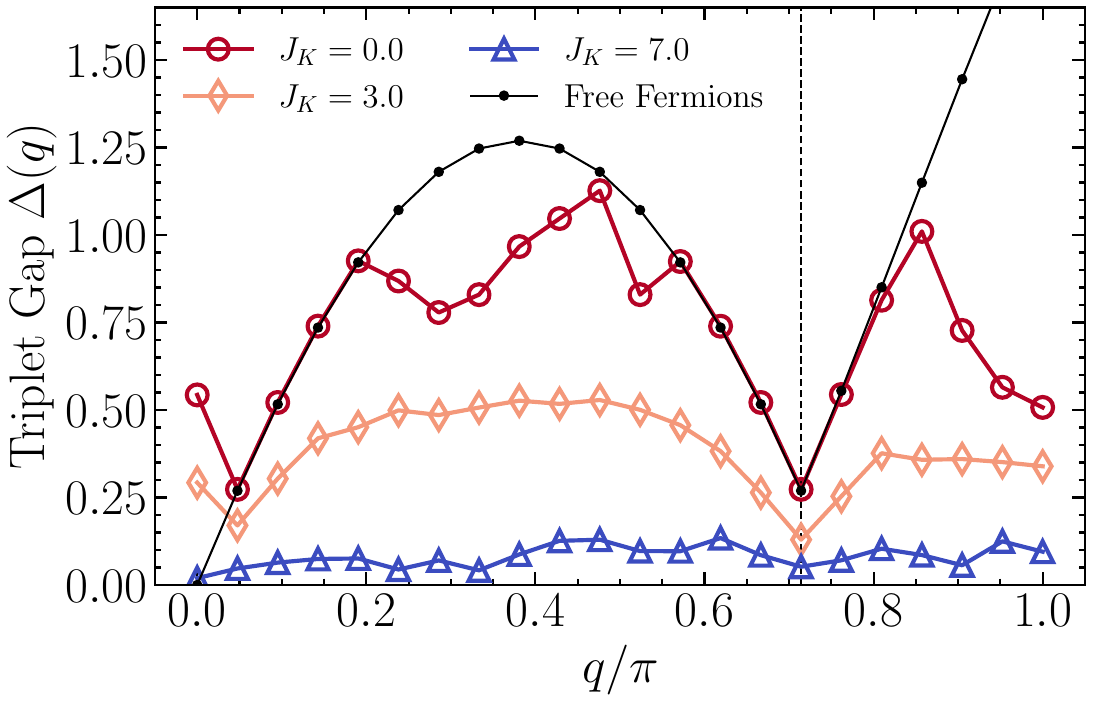}}
        \caption{\label{fig:L_42_momentum} Triplet gap at different momenta $q$ for values of $t=1.0$, $J_1=J_2=J_\perp=1.0$ at a filling $\delta \approx 0.2857$ on a chain of $N=42$ sites. The red circles, orange rhombi, and blue triangles correspond to a decoupled $(J_K=0.0)$, Kondo-unscreened $(J_K=3.0)$, and Kondo-screened $(J_K=7.0)$ phases, respectively. Black points correspond to single-particle excitations of a one-dimensional free fermion system, computed analytically. Statistical error bars on the gaps are smaller than the symbol size. Vertical dashed line is at momenta $q=2k_F$.}
    \end{center}
\end{figure}

Finally, it is worth noting that the phase diagram of the ALM model can be viewed as the reverse of that of the Kondo–Heisenberg model. In the Kondo–Heisenberg model, the system is described by the LL$^*$ phase (unstable to Luther-Emery phase) at small Kondo coupling and Luttinger Liquid at larger Kondo coupling~\cite{Nikolaenko2024}. This contrast helps explain the success of the ALM model in capturing the phase diagram of cuprate superconductors in two dimensions. Indeed, the FL phase naturally maps to the overdoped regime, while the FL$^*$ explains more intricate properties of the pseudogap regime. Such a correspondence is difficult to realize in a two-layer model, where the FL$^*$ phase would instead have to describe a conventional overdoped regime, which is quite inconsistent with experimental observations.

\subsection{Momentum Dispersion} \label{sec:momentum}
A key advantage of periodic boundary conditions is the restoration of translational invariance, which enables variational optimization directly within Hilbert-space subsectors with definite momentum $q$.
In this setting, the Transformer-based Ansatz can efficiently access the triplet excitation spectrum at finite momentum. We define the triplet gap as $\Delta(q)=E_{\text{triplet}}(q)-E_0$, where $E_0$ is the ground state energy and $E_{\text{triplet}}(q)$ is estimated by performing Monte Carlo sampling in the sector $S^z=1$~\cite{viteritti2025prb}, with a fixed value of momentum $q$ enforced by symmetry-projection operators~\cite{nomura2021iop, viteritti2022,zhang2025neuraltransformerbackflowsolving}. Fig.~\ref{fig:L_42_momentum} shows the triplet gap for three representative values of $J_K$ corresponding to the decoupled, Kondo-unscreened, and Kondo-screened phases. For this analysis, we set $J_1=J_2=J_\perp=1.0$ to make the spin gap in the decoupled phase larger. In this regime, the transition from Luttinger-liquid behavior to the Kondo-screened regime occurs around $J_K \approx 5.0$; refer to \cref{app:transition} of the \textit{Appendix}.

For the decoupled case ($J_K=0.0$), the physical layer reduces to a free fermion system, so its triplet excitations should coincide with the single-particle particle-hole continuum, which we compute analytically (see black points in Fig.~\ref{fig:L_42_momentum}). At low energies the numerical gaps follow the analytical continuum very closely, with a linear dispersion emerging near $q=0$ and near $q=2k_F$, as expected. At higher energies, however, the numerical spectrum departs from the free-fermion prediction: even at $J_K=0.0$ the Ansatz retains the ancilla layers, which are gapped by the intra-ancilla coupling $J_\perp$. This spin gap sets an additional energy scale that is absent in the pure free-fermion problem.

In the regime of small Kondo coupling ($J_K=3.0$), the triplet spectrum still has a pronounced minimum at $q=2k_F$, in agreement with the Luttinger-liquid behavior. The overall magnitude of the triplet gap is reduced compared to the decoupled case, which implies that the bandwidth of the system becomes smaller. The slope of the dispersion near Fermi points decreases; therefore, the velocity of the excitations decreases.

\begin{figure}[t]
    \begin{center}
\centerline{\includegraphics[width=\columnwidth]{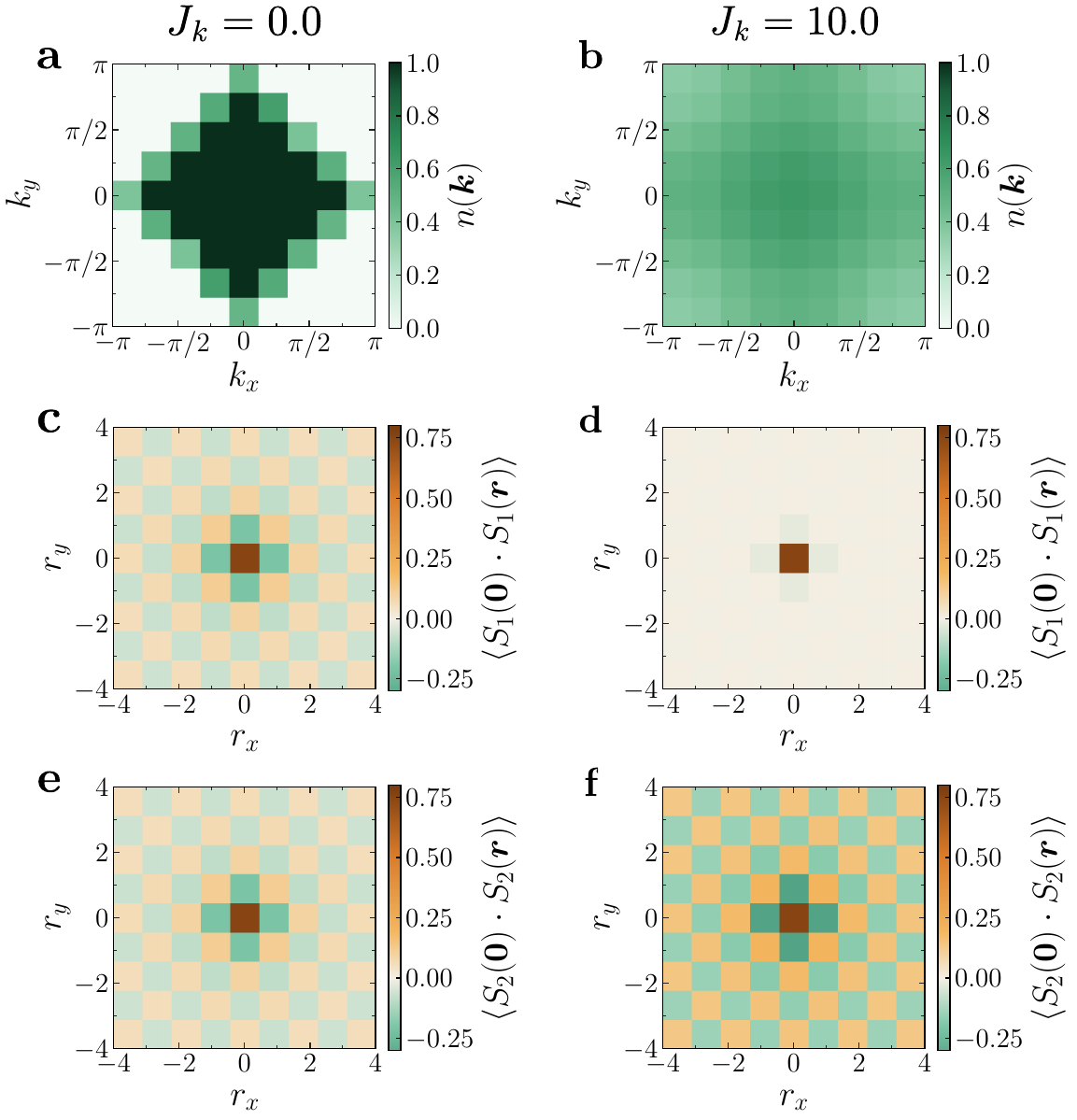}}
\caption{\label{fig:2d_benchmark} Two-dimensional benchmark of the Transformer Ansatz on the ALM on an $8\times 8$ square lattice with periodic boundary conditions, at half-filling and $t=1$, $J_1=J_2=0.5$, $J_\perp=1.25$, for the two limiting Kondo couplings $J_K=0$ (left) and $J_K=10$ (right). Top row (a,b): electronic momentum distribution $n(\mathbf{k})$. Middle row (c,d): first-ancilla spin correlations $\langle \mathbf{S}_1(0)\cdot\mathbf{S}_1(\mathbf{r})\rangle$. Bottom row (e,f): second-ancilla spin correlations $\langle \mathbf{S}_2(0)\cdot\mathbf{S}_2(\mathbf{r})\rangle$.}
    \end{center}
\end{figure}

In the Kondo-screened phase at $J_K=7.0$, we observe an even more drastic suppression of the amplitude of the triplet gap, followed by a significant reduction in the quasi-particle velocity. The phenomenon is widely observed in heavy-fermion physics: as the conduction electrons and localized spins mix, the renormalized bands become much narrower, and the effective mass increases~\cite{Coleman2006,Danu2021,Boulder25}. Overall, as the Transformer architecture is well-suited for periodic boundary conditions, it provides momentum-resolved information about the excitation spectrum and offers a clearer understanding of the physics of Kondo screening.

\section{Benchmark in Two Dimensions}\label{sec:benchmark2d}

We now test whether the same architecture transfers to the two-dimensional ALM, the setting of interest for the cuprates, where controlled numerical benchmarks are scarce. Mapping the full phase diagram is beyond the scope of this work, but we verify that the Transformer Ansatz reproduces the two analytically understood limits: the decoupled regime $J_K=0$ and the strong-coupling Kondo-screened regime $J_K\to\infty$. We use an $8\times8$ square lattice with PBC at half filling ($\delta=0$, $N_e=64$), $t=1$, $J_1=J_2=0.5$, fixing $J_\perp=1.25$ in order to be able to compare with QMC results for the two-layer Heisenberg model from Ref.~\cite{qmc2dbenchmark}. The same architecture and optimization setup as in the one-dimensional case (refer to \cref{sec:architecture} of the \textit{Appendix}) are used. \cref{fig:2d_benchmark} reports the momentum distribution $n(\boldsymbol{k})=\frac{1}{2}\sum_\sigma\langle\hat{c}^\dagger_{\boldsymbol{k}\sigma}\hat{c}_{\boldsymbol{k}\sigma}\rangle$ (top) and the ancilla spin correlations $\langle\mathbf{S}_{1,2}(0)\cdot\mathbf{S}_{1,2}(\mathbf{r})\rangle$ (middle, bottom) for $J_K=0$ (left) and $J_K=10$ (right).

At $J_K=0$ the fermions decouple from the ancilla sector. The momentum distribution [see \cref{fig:2d_benchmark}(a)] shows the sharp step of a \emph{small} Fermi surface set by the electron density alone, as required by Luttinger's theorem, while the two ancilla layers reduce to the bilayer Heisenberg model. As a stringent check, we compute the intraplane staggered structure factor ${S_{\rm intra} = \frac{1}{L^2}\sum_{i,j}\langle S^z_{2,i} S^z_{2,j}\rangle (-1)^{i-j}}$ and the interplane one $S_{\rm inter} = \frac{1}{2L^2}\sum_{i,j}\langle (S^z_{1,i}-S^z_{2,i})(S^z_{1,j}-S^z_{2,j})\rangle (-1)^{i-j}$, which are directly related to the single-plane and full staggered magnetizations. We obtain $S_{\rm intra} = 5.95(1)$ and $S_{\rm inter} = 11.5(1)$, in close agreement with the quantum Monte Carlo results for the two-layer Heisenberg model at the same interlayer coupling, $5.97$ and $11.9$~\cite{qmc2dbenchmark}. This is the two-dimensional analog of the decoupled LL phase and, in the cuprate mapping, of the overdoped Fermi liquid.

At $J_K=10$ the first ancilla layer is Kondo-screened and absorbed into a coherent heavy band, reconstructing the Fermi surface from small to \emph{large}: its volume now counts one localized spin per site in addition to the electrons, the two-dimensional counterpart of the $2k_F\!\to\!2k_F^*$ shift seen in the one-dimensional case~\cite{Senthil2003}. The screening is visible in the first-layer correlations [see \cref{fig:2d_benchmark}(d)], which are strongly suppressed beyond the on-site value, while the second layer, freed from the rung singlet, orders antiferromagnetically in a $(\pi,\pi)$ N\'eel pattern [refer to \cref{fig:2d_benchmark}(f)]. As a second check, the corresponding intraplane structure factor $S_{\rm intra} = 11.3(1)$ is close to the value $11.37$ obtained from quantum Monte Carlo for the isolated square-lattice Heisenberg model~\cite{qmcheisenbergbenchmark}. That a single unmodified Ansatz captures both the small Fermi surface at $J_K=0$ and the enlarged volume, local screening, and N\'eel order at $J_K=10$ demonstrates that the construction transfers to the physically relevant two-dimensional regime, where the periodic $8\times8$ geometry places these results out of reach for DMRG.

\section{Conclusions}\label{sec:conclusions}
We have introduced a Transformer-based NQS for lattice models with composite local Hilbert spaces, combining fermionic and spin degrees of freedom within a unified variational framework. By explicitly tokenizing the local Hilbert space, following a procedure similar to the one used in natural language processing~\cite{vaswani2017attention,jm3}, the proposed Ansatz naturally incorporates heterogeneous on-site structures and captures correlations both within a site and across the lattice.

Applying this approach to the one-dimensional Ancilla Layer Model, we demonstrated excellent quantitative agreement with DMRG for ground-state energies and correlation functions. The Transformer wave function accurately reproduces momentum-resolved charge and spin correlations across a broad range of Kondo couplings and provides a faithful description of both weakly coupled Luttinger-liquid behavior (LL) and Kondo-screened regimes (LL*). By combining our variational results with DMRG calculations, we further characterized the phase diagram of the model and related it to the evolution of correlation functions.

In the context of two-dimensional ALM models of the cuprate phase diagram, the LL phase is the analog of the Fermi liquid phase at large doping, while the LL* phase is the analog of the fractionalized Fermi liquid (FL*) phase applied to the pseudogap metal at low doping. We also found an instability of the LL* phase to a Luther-Emery (LE) phase, and this can be viewed as a one-dimensional prototype of the onset of superconductivity from the pseudogap in the cuprates.

A key advantage of the NQS approach is its ability to perform similarly for both open and periodic boundary conditions. In particular, the rapid convergence of the energy with system size under periodic boundary conditions highlights the potential of this method for accessing bulk properties without incurring the computational overhead typically associated with tensor-network methods. This feature is expected to become especially important in two-dimensional lattice systems, where boundary effects are more severe and periodic boundary conditions are often essential for reliable thermodynamic-limit estimates~\cite{viteritti2025prb}. Building on this, we took a first step into two dimensions, benchmarking the method on an $8\times8$ square lattice with periodic boundary conditions. Without any modification, the same architecture reproduces the two analytically understood limits of the model, the small Fermi surface at $J_K=0$ and the Kondo-screened large Fermi surface with N\'eel order at large $J_K$, with ancilla spin correlations in quantitative agreement with quantum Monte Carlo. This confirms that the construction transfers to the physically relevant two-dimensional regime, in a geometry out of reach for DMRG.

More broadly, our results demonstrate that this architecture provides a flexible and accurate variational representation for correlated systems with composite local Hilbert spaces. The token-based formulation introduced here offers a systematic route toward the study of higher-dimensional models, multi-orbital systems, and other strongly correlated settings in which conventional approaches face significant challenges.

\begin{figure*}[t]
    \begin{center}
\centerline{\includegraphics[width=2\columnwidth]{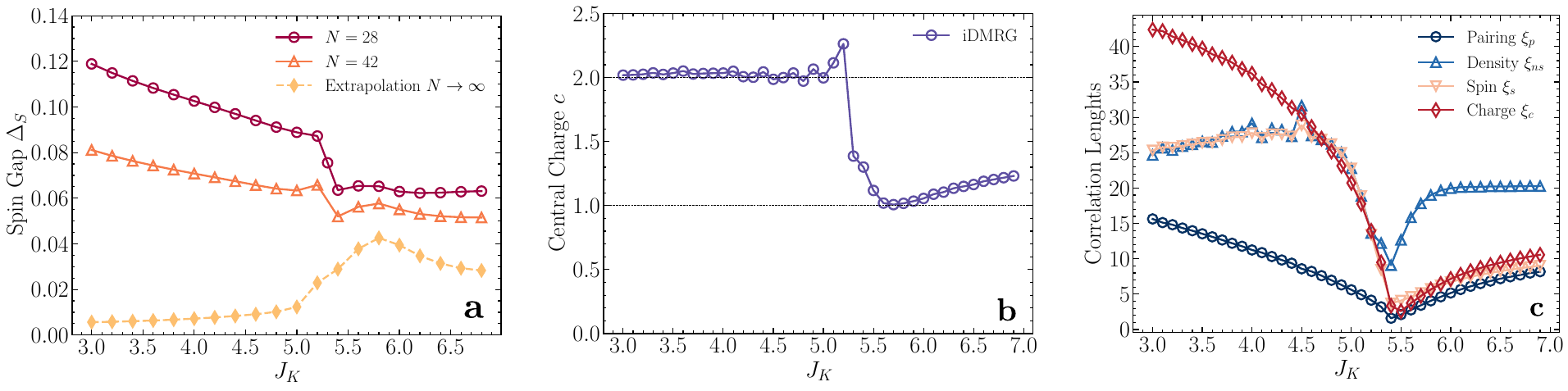}}
        \caption{\label{fig:dmrg} \textbf{Panel a:} The spin gap $\Delta_S$ as a function of the Kondo coupling $J_K$, obtained with finite DMRG with bond dimension $\chi=1000$ for chains of length $N=28$ (purple curve), $N=42$ (orange curve) and the extrapolation to $N \to \infty$ (yellow curve). \textbf{Panel b:} The central charge as a function of the Kondo coupling $J_K$, obtained with iDMRG with bond dimensions $\chi=500$ and $1000$. \textbf{Panel c:} Spin ($\xi_s$), charge ($\xi_c$), density ($\xi_{ns})$, and pairing ($\xi_p$) correlation lengths as a function of the Kondo coupling $J_K$, obtained with iDMRG for $\chi=1000$. Parameters are  $t=1.0$, $J_1=J_2=J_\perp=1.0$ at a filling $\delta \approx 0.2857$.}
    \end{center}
\end{figure*}

\section{Appendix}
\subsection{NQS Architecture and Optimization Details}\label{sec:architecture}
For most of the calculations in this work, we use a Transformer with $n_l = 4$ layers, $h = 12$ attention heads, and an embedding dimension of $d = 72$ (see Ref.~\cite{viteritti2025prb} for more details about their role). For a system size of $N=42$, this choice results in a total of $P\approx 4 \times 10^5$ variational parameters. The Transformer state is optimized within the Variational Monte Carlo (VMC) framework~\cite{becca2017} using Stochastic Reconfiguration (SR)~\cite{Sorella1998} with the linear algebra trick~\cite{rende2024stochastic,chen2024empowering}, enhanced via the MARCH optimizer~\cite{gu2025solvinghubbardmodelneural, spring}. 
The optimization is carried out for $10^4$ steps with $M=2^{13}$ Monte Carlo samples per iteration. The learning rate is initialized to $\eta=0.01$ and is gradually annealed during the optimization.

During the VMC sampling, configurations of the variational wave function are generated by proposing, with equal probability, three types of Monte Carlo updates: fermionic hopping moves between nearest neighbor sites, spin exchanges between nearest neighbors, and spin–fermion exchange moves that couple the fermionic degrees of freedom to one of the spin chains.

\subsection{Decoupled limit $J_K=0.0$}
An important conceptual aspect of the present wave function construction concerns the limit in which the electronic and spin sectors are independent, and the exact wave function factorizes into a product of a fermionic and a spin component ($J_K=0.0$ for the ALM). In our architecture, no explicit factorization between these degrees of freedom is imposed: the Transformer representation is always used to predict a set of backflow orbitals. Despite this unified parametrization, we observe that the Ansatz performs remarkably well in the decoupled regime, indicating that the network can effectively learn the appropriate factorized structure when required. While one could, in principle, enforce a separate spin factor multiplied by a Slater determinant, our results suggest that allowing the model to process all local degrees of freedom jointly leads to a more flexible and robust optimization across the different coupling regimes.

\subsection{DMRG Computational Details}
The DMRG calculation was performed using the TeNPy library (version 0.10.0)~\cite{Tenpy2018}. The bond dimension $\chi$ varied from $500$ to $1000$ with the number of sweeps $n_{\text{sweeps}}=20$. The maximum discarded weight was below $10^{-5}$ and the energy converged to $10^{-8}$ between the sweeps. iDMRG simulations were performed using a translationally invariant iMPS ansatz with unit cell size $L=14 $, with the bond dimension $\chi$ up to $2000$ and with the maximum discarded weight below $10^{-5}$.

\subsection{Nature of the Phase Transition}
\label{app:transition}

In this Section, we elaborate on the transition between the Kondo-unscreened LL phase with Fermi momentum $k_F$ and the Kondo-screened phase with Fermi momentum $k_F^{\ast}$. We use parameters $t=1.0$, $J_1=J_2=J_\perp=1.0$ as in \cref{sec:momentum}.  Fig.~\ref{fig:dmrg}(a) shows the spin gap ${\Delta_S=E(S_z=1) - E(S_z=0)}$ as a function of the Kondo coupling. In the Kondo-screened phase, the gap becomes non-zero, suggesting that for large $J_\perp$, the LL$^*$ is ultimately unstable toward the Luther-Emery phase. A similar conclusion was drawn in Refs. \cite{Nikolaenko2024,Berg2010,Zhang2022}. However, for small $J_\perp$, a true LL$^*$ phase is realizable, as shown in Fig.~\ref{fig:phase_diagram}. This contrasts ALM with the Kondo-Heisenberg model~\cite{Nikolaenko2024}, in which the LL$^*$ phase existed only for ferromagnetic exchange interactions.

Fig.~\ref{fig:dmrg}(b) shows the central charge, extracted via iDMRG~\cite{mcculloch2008} using the entanglement entropy of a half-infinite chain $S=\frac{c}{6} \ln[\xi(\chi)]$, where $\xi(\chi)$ is the largest correlation length at a  fixed bond dimension $\chi$~\cite{Pollmann2009,Calabrese2004}. At $J_K<J_{Kc}$, the central charge $c\approx2$, while at $J_K>J_{Kc}$ it changes to $c\approx1$. This confirms our observation that the $k_F$ Fermi momentum phase is a C1S1 Luttinger liquid and the $k_F^\ast$ Fermi momentum phase is C1S0 Luther-Emery phase.

Finally, Fig.~\ref{fig:dmrg}(c) shows the correlation lengths across the transition for a finite bond dimension $\chi$. The density correlation length ($\xi_{ns}$) is larger than the pairing correlation length ($\xi_p$), which means that the phase is unstable toward CDW order. In the region $J_K\in [5.3,5.6]$, the correlation lengths become extremely short, on the order of several lattice spacings, which implies the ultra-local nature of the quantum critical point, thoroughly discussed in Ref.~\cite{Nikolaenko2024}.

\begin{acknowledgments}
We thank G. Carleo for useful discussions. AN and SS are supported by the U.S. National Science Foundation grant No. DMR-2245246 and by the Simons Collaboration on Ultra-Quantum Matter which is a grant from the Simons Foundation (651440, S.S.). This work was supported as part of the ``Swiss AI initiative'' by a grant from the Swiss National Supercomputing Centre (CSCS) under project ID a117 on Alps. RR and LLV acknowledge the CINECA award under the ISCRA initiative for the availability of high-performance computing resources and support. RR acknowledges support from the Flatiron Institute. The Flatiron Institute is a division of the Simons Foundation. YHZ is supported by the Alfred P. Sloan Foundation through a Sloan Research Fellowship. LLV is supported
by SEFRI under Grant No. MB22.00051 (NEQS - Neural Quantum).
\end{acknowledgments}

\bibliography{refs}

\end{document}